\documentclass[aps,physrev,twocolumn,superscriptaddress,nofootinbib]{revtex4-2}
\usepackage{graphicx}
\usepackage{longtable}
\usepackage{amsmath}
\usepackage[overload]{textcase}

\begin{document}

\title{Itinerant antiferromagnetism in the antagonistic pair compound Y$_4$Co$_3$Ag}

\author{Rafaela F. S. Penacchio}
\affiliation{Ames National Laboratory, US DOE, Iowa State University, Ames, IA, USA}
\affiliation{Department of Physics and Astronomy, Iowa State University, Ames, Iowa 50011, USA}
\affiliation{Institute of Physics, University of S{\~{a}}o Paulo, S{\~{a}}o Paulo, SP, Brazil}

\author{Nao Furukawa}
\affiliation{Ames National Laboratory, US DOE, Iowa State University, Ames, IA, USA}
\affiliation{Department of Physics and Astronomy, Iowa State University, Ames, Iowa 50011, USA}

\author{Joanna M. Blawat}
\affiliation{National High Magnetic Field Laboratory,
Los Alamos National Laboratory, Los Alamos, NM, USA.}

\author{John Singleton}
\affiliation{National High Magnetic Field Laboratory,
Los Alamos National Laboratory, Los Alamos, NM, USA.}

\author{Zhouqi Li}
\affiliation{Ames National Laboratory, US DOE, Iowa State University, Ames, IA, USA}
\affiliation{Department of Physics and Astronomy, Iowa State University, Ames, Iowa 50011, USA}

\author{Raquel~A. Ribeiro}
\affiliation{Ames National Laboratory, US DOE, Iowa State University, Ames, IA, USA}
\affiliation{Department of Physics and Astronomy, Iowa State University, Ames, Iowa 50011, USA}

\author{S{\'{e}}rgio L. Morelh{\~{a}}o}
\affiliation{Institute of Physics, University of S{\~{a}}o Paulo, S{\~{a}}o Paulo, SP, Brazil}

\author{Sergey L. Bud’ko}
\affiliation{Ames National Laboratory, US DOE, Iowa State University, Ames, IA, USA}
\affiliation{Department of Physics and Astronomy, Iowa State University, Ames, Iowa 50011, USA}

\author{Paul C. Canfield}
\affiliation{Ames National Laboratory, US DOE, Iowa State University, Ames, IA, USA}
\affiliation{Department of Physics and Astronomy, Iowa State University, Ames, Iowa 50011, USA}

\author{Tyler J. Slade}
\affiliation{Ames National Laboratory, US DOE, Iowa State University, Ames, IA, USA}
\affiliation{Department of Physics and Astronomy, Iowa State University, Ames, Iowa 50011, USA}

\date{\today}

\begin{abstract}

Low dimensional crystallographic motifs have long been associated with desirable physical properties. The confinement of electrons to low dimensions is thought to enhance quantum fluctuations and may promote correlated phenomena. Here, using the antagonistic pair concept, we add Y to the immiscible Co-Ag pair to discover Y$_4$Co$_3$Ag. This compound adopts a monoclinic $I$2/$m$ structure consisting of Y channels that are filled by one-dimensional zigzag and hexagonal Co chains, which extend along the crystallographic $b$-axis with no nearest neighbor contacts between Co and Ag atoms. Transport, magnetic, and specific heat measurements reveal that Y$_4$Co$_3$Ag orders antiferromagnetically at $T_N=14.9$\,K with an effective magnetic moment $\mu_{\text{eff}}$ = 1.4 $\mu_{\text{B}}$/Co.  Specific heat measurements show only a small entropy loss on the order of $0.1\,R\ln2$ associated with magnetic order, and magnetization isotherms, in DC fields up to 70\,kOe at 1.8\,K and in pulsed fields up to 600\,kOe at 500\,mK, indicate a small ordered moment of less than 0.2 \,$\mu_B$/Co. Taken together, our results imply the presence of small, itinerant moments and strong fluctuations in Y$_4$Co$_3$Ag, suggesting that Y$_4$Co$_3$Ag may be a promising candidate material to investigate itinerant magnetic interactions in a quasi-one dimensional system.
\end{abstract}

\maketitle

\textit{Introduction.} Materials with low dimensional crystal structures can exhibit enhanced fluctuations that may promote emergent phases with exotic behavior \cite{sachdev2000, mermin1966}. Prominent examples include unconventional superconductivity, charge density waves, and quantum Hall effects \cite{stewart2017unconventional,plakida2010high,stewart2011superconductivity,jerome2024quasi,von202040}. Low dimensional spin systems also can produce a wealth of unconventional ground-states and excitations, such as spin-charge separation \cite{auslaender2005spin,kim1996observation}, Bose-Einstein condensation of integer spin quasiparticles \cite{giamarchi2008bose,aczel2009field}, and spin liquid physics \cite{broholm2020quantum,takagi2019concept}. Likewise, specific low-dimensional arrangements, such as kagome or square nets support topologically protected electronic states \cite{checkelsky2024flat,kang2020topological,han2021evidence,neves2024crystal, klemenz2019topological,klemenz2020role}. This relation between reduced dimensionality and emergent physical phenomena strongly motivates the continued discovery and investigation of compounds in which atoms, especially those bearing magnetic moments, adopt low-dimensional crystallographic motifs.

The antagonistic pairs concept is an emerging strategy to design intermetallic compounds containing low-dimensional structural units \cite{canfield2019new, slade2024}, where the challenge is to introduce a suitable third element \textit{C} to a strongly immiscible pair of elements \textit{A} and \textit{B}, such that an ordered ternary compound $A_{\text{x}}B_{\text{y}}C_{\text{}z}$ can be formed. We have found that when such ternary phases do exist, the strong chemical incompatibility between \textit{A} and \textit{B} is reflected in the ternary crystal structure, with \textit{C} spatially separating the immiscible elements into distinct, often low-dimensional substructures such as sheets, chains, or clusters. In this way, the antagonistic pairs premise represents a generalizable route for discovering ternary compounds with quasi-low-dimensional motifs intrinsically built into the crystal structure.  

Motivated by this concept, we report in this Letter the discovery and physical properties of Y$_4$Co$_3$Ag. This is the first Y--Co--Ag ternary compound and, owing to the strong Co--Ag immiscibility, Y$_4$Co$_3$Ag has a remarkable monoclinic (\textit{I}2/\textit{m}) crystal structure with zigzag and hexagonal Co chains that run down the \textit{b}-axis. The Co chains are fully encapsulated inside of Y channels, such that there are no nearest neighbor Ag--Co contacts. Transport, magnetic, and specific heat measurements reveal antiferromagnetic order at $T_N$ = 14.9\,K with a small entropy loss on the order of 0.1$R\ln{2}$. Anisotropic magnetic susceptibility yields paramagnetic (Curie-Weiss) effective moments with magnitude of $\mu_{eff}=$\,1.4\,$\mu_B$/Co, and Weiss temperatures ranging from $-107(1)$\,K to $-130(4)$\,K. Pulsed field measurements reveal that Y$_4$Co$_3$Ag does not reach saturation in fields up to 600\,kOe, but the magnetization has modest values consistent with a small ordered moment, likely under 0.2 $\mu_B$/Co. Taken together, our measurements indicate that the one-dimensional Co chains in Y$_4$Co$_3$Ag support itinerant, small moment, antiferromagnetism, with magnetic fluctuations likely persisting well above $T_N$. Therefore, this work demonstrates the successful use of the antagonistic pair concept for designing compounds with magnetic ions embedded in low-dimensional substructures and establishes Y$_4$Co$_3$Ag as a potential platform to explore quasi-1D magnetic interactions and ordering in the itinerant magnetism regime.

\textit{Methods.} Whereas Co and Ag are strongly immiscible at temperatures up to at least 1700\,$^\circ$C, both elements readily mix with Y \cite{okamotoCoY, okamotoAgY}. In particular, the Y--Co phase diagram exhibits a relatively deep eutectic near $\approx 60\%$ Y, allowing access to a single-phase liquid below 1000\,$^\circ$C for $\approx50-75\%$ Y. This suggests that Y might be an attractive third element for growing ternary compounds containing both Co and Ag. These considerations are similar to those that led to the discovery of La$_4$Co$_4$X (X = Pb, Bi) \cite{slade2024}, where we explored compositions near to the Co--La eutectic to grow compounds containing the antagonistic pairs Co--Pb and Co--Bi. Following this approach, single crystals of Y$_4$Co$_3$Ag were grown by adding small quantities of Ag to Y-Co melts as follows. 

\begin{figure}[t]
    \centering
    \includegraphics[scale=1]{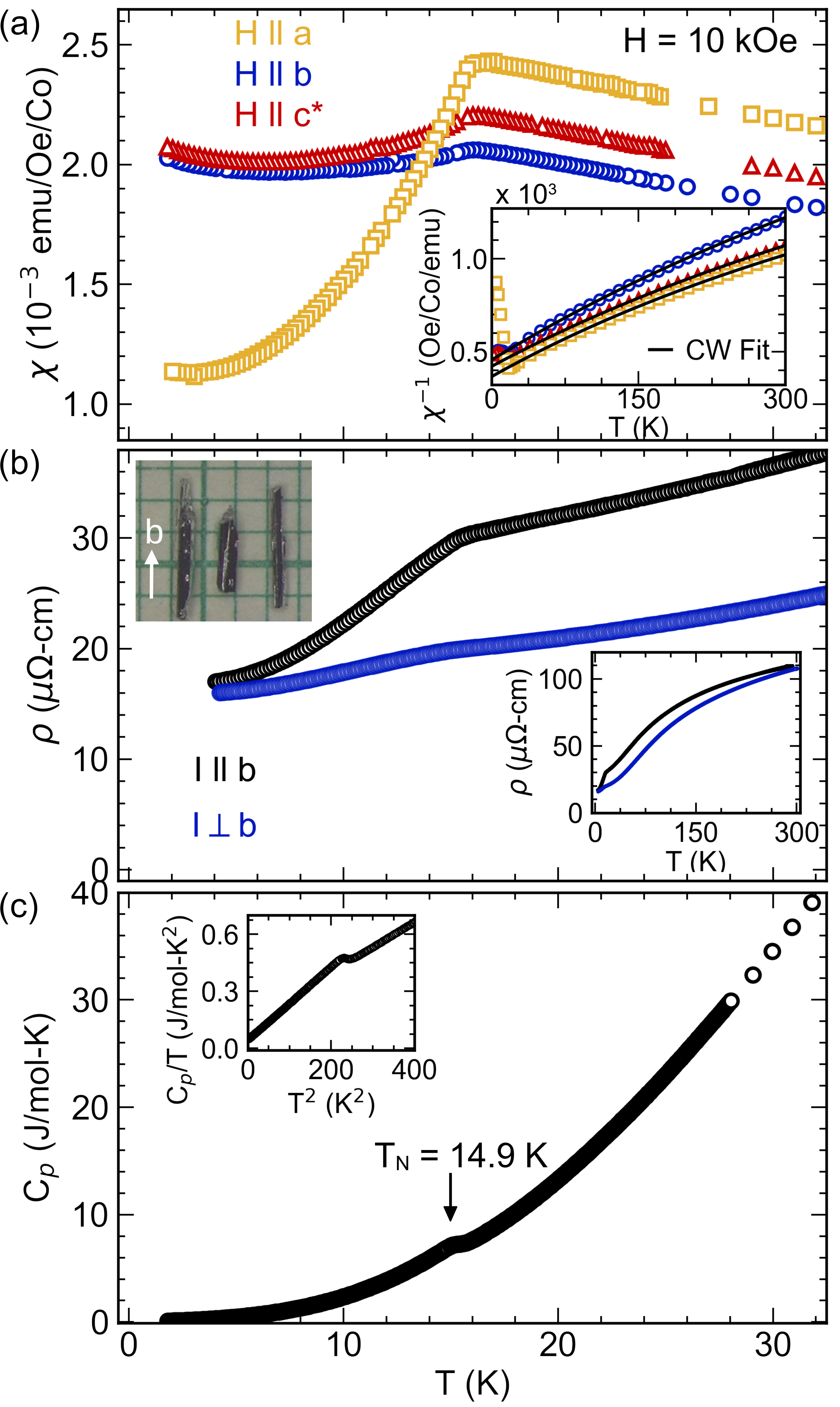}
    \caption{Low-temperature physical properties of Y$_4$Co$_3$Ag. (a) Anisotropic temperature-dependent magnetic susceptibility $\chi=M/H$. The inset shows $\chi^{-1}$, and the solid curves are the Curie-Weiss fits (see text) to the data. (b) Temperature dependence of the electrical resistivity $\rho$ measured with current applied parallel (black) and perpendicular (blue) to the $b$-axis. The right inset shows $\rho(T)$ for $T$ = 1.8--300\,K, and the left inset shows a picture of typical rod-like Y$_4$Co$_3$Ag crystals on a mm-scale grid. (c) Temperature dependence of the molar heat capacity, $C_p$, for temperatures $1.8\leq T \leq 32$\,K. The inset shows $C_p/T$ against $T^2$, showing linear dependence below $T_N$, and yielding $\gamma = 50.1(5)$\,mJ/mol-K$^2$, $\beta\sim1.5$\,mJ/mol-K$^4$, and Debye temperature $\Theta_D = 108.1(2)$\,K.}
    \label{fig:properties}
\end{figure}

Elemental Y (Ames Laboratory, 99.9+\%), Co pieces (American Elements, 99.95\%), and Ag (DOE stockpile, 99.9\%) were weighted in a molar ratio of Y$_{60}$Co$_{40}$Ag$_{10}$ and melted together using a home-built arc furnace under partial Ar atmosphere. The arc melting step was used to ensure intimate mixing between the individually high-melting elements Y and Co, allowing the mixture to readily melt into a single phase liquid below 1200\,$^\circ$C. The melted button was then placed in a Ta crucible set with homemade Ta caps and filter \cite{canfield2019, canfield2001}. The Ta crucible was sealed under an Ar atmosphere using the arc melter, and then flame-sealed into a fused silica ampule that was backfilled with $\approx 1/6$ atm Ar gas. The ampule was heated in a box furnace to 1150\,$^\circ$C and held at that temperature for 6\,h. Then, the furnace was gradually cooled to 760\,$^\circ$C over 125\,h, after which the tube was removed from the furnace and the excess flux was decanted in a centrifuge with metal cups and rotors \cite{canfield2001}. After cooling to room temperature, the tube was opened to reveal a mixture of rod- and block-like crystals. The left inset of Fig. \ref{fig:properties}(b) shows typical examples of the rod-like crystals. Elemental analysis and single crystal X-ray diffraction indicated both morphologies to be the ternary Y$_4$Co$_3$Ag. We also succeeded in growing Y$_4$Co$_3$Ag single crystals using fritted Al$_2$O$_3$ crucibles \cite{canfield2016, LSP}. Because elemental analysis and powder X-ray diffraction indicated a small Al contamination in samples grown in Al$_2$O$_3$, all the discussions in the main text are based on single crystals grown in Ta tubes. Comparison to the Al$_2$O$_3$ grown samples is presented in the Supporting Information (SI). 

The chemical composition of the samples was determined by Energy Dispersive Spectroscopy (EDS) quantitative chemical analysis using a ThermoFisher Teneo Lovac field-emission scanning electron microscope (FE-SEM). The data was analyzed using an Oxford Instruments Aztec System with an X-Max-80 detector, attached to Teneo. The measurements were conducted with an acceleration voltage of 20\,kV, current of 1.6 nA, working distance of 10 mm, and a takeoff angle of 35$^\circ$. The standards used for reference are internal to the OXFORD software. 

Single crystal X-ray diffraction was performed using a Rigaku XtaLab Synergy-S diffractometer with Ag radiation ($\lambda = 0.56087$\,\AA), in transmission mode, operating at 65 kV and 0.67 mA. The samples were held in a nylon loop with Apiezon N grease, and the data was collected at room temperature. The total number of collected runs and images was based on the strategy calculation from CrysAlisPro (Rigaku OD,
2023). Data integration and reduction were also performed using CrysAlisPro, and a numerical absorption correction was applied based on Gaussian integration over a multifaceted crystal model. The structures were solved by intrinsic phasing using the SHELXT software package and were refined with SHELXL. 

Powder X-ray diffraction patterns were obtained using a Rigaku Miniflex-II instrument operating with Cu-$K\alpha$ radiation with $\lambda=1.5406\,$\AA\, ($K\alpha_1$) and $\lambda=1.5443\,$\AA\, ($K\alpha_2$) at 30\,kV and 15 mA. As Y$_4$Co$_3$Ag grown 
in Ta tubes had two characteristic morphologies, the samples were prepared by grinding a representative number of crystals of each morphology into a fine powder. The powder patterns were refined using the Rietveld method with GSAS-II software \cite{toby2013}.

\begin{figure*}[ht]
    \centering
    \includegraphics[scale=1.1]{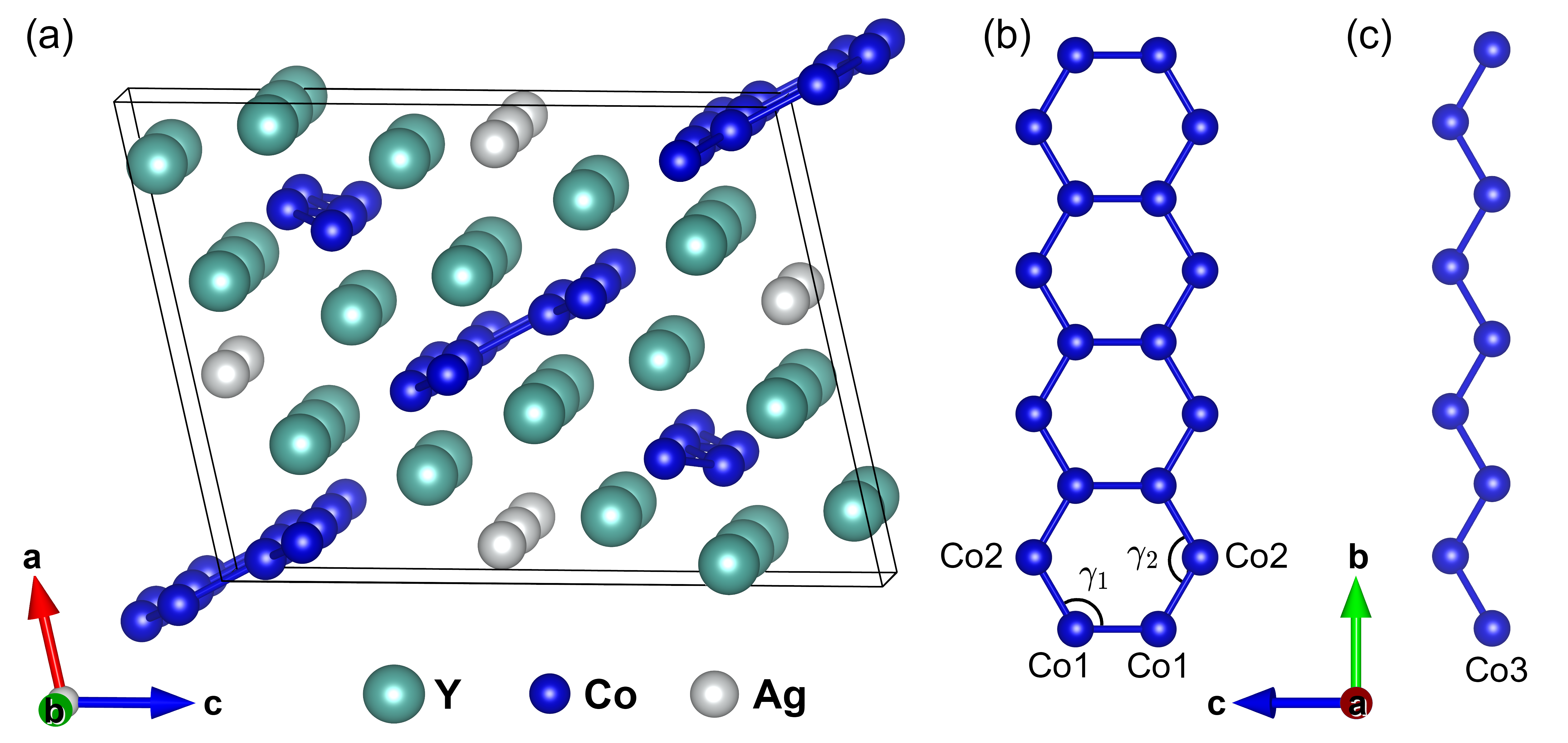}
    \caption{Crystal structure of Y$_4$Co$_3$Ag. (a) A view of the crystal structure along the $b$-axis, emphasizing the complete sheathing of Co by Y atoms with no nearest-neighbor Co-Ag contacts. Both Y and Ag form linear chains running along the $b$-axis of the monoclinic ($I$2/$m$) unit cell. To highlight these chains, (a) shows a projection of two unit cells along the (010) direction. The intrachain distance between two Y or two Ag atoms is 3.9197(3) \AA, i.e., the unit cell length in the (010) direction. The Co ions form both (b) hexagonal and (c) zigzag chains that occupy the channels formed by Y. Co-Co distances within the hexagonal chains are $d_{\text{Co1-Co1}} = 2.3432(6)$ \AA\,and $d_{\text{Co1-Co2}} = 2.2893(6)$ \AA, yielding internal angles of $\gamma_1 = 121.12(1)^\circ$ and $\gamma_2 = 117.76(2)^\circ$. The nearest-neighbor distance within the Co zigzag chain is 2.3327(6) \AA. The minimum intra-chain Co-Co distance is $\sim 4.2$\,\AA.} 
    \label{fig:structure}
\end{figure*}

Temperature-dependent resistance measurements were performed between 1.8 and 300\,K within a Quantum Design Physical Properties Measurement System (PPMS). The AC resistance was measured using a LakeShore AC resistance bridge (model 372), with a frequency of 16.2 Hz and a 3.16 mA excitation current. The resistance measurements were performed with current applied parallel or perpendicular to the crystallographic $b$-axis, i.e., the long axis of the rod-like samples. Contacts were made by spot welding a 25\,$\mu$m thick annealed Pt wire onto the faces of the crystals in the standard four-point geometry. After spot welding, a small amount of silver epoxy was painted onto the contacts to ensure good mechanical strength, yielding typical contact resistances of $\approx  1\,\Omega$.

\begin{figure}[t]
    \centering
    \includegraphics[scale=1]{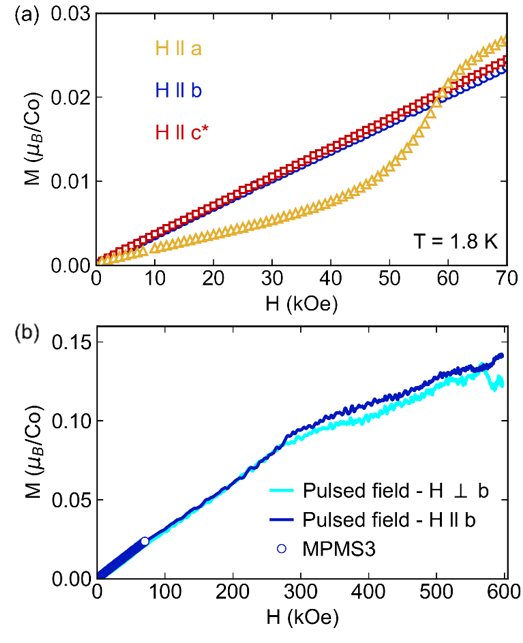}
    \caption{Field-dependent magnetization isotherms for Y$_4$Co$_3$Ag measured at (a) $T=1.8$\,K in a DC field and (b) 500\,mK (solid curves) in a pulsed field. The points (open circles) in (b) show the low-field data used to scale the pulsed-field data (see \textit{Methods}).}
    \label{fig:magnetic}
\end{figure}

Low-field (0--70\,kOe) magnetization measurements were performed in a Quantum Design Magnetic Property System (MPMS3) SQUID magnetometer operating in the DC measurement mode. The measurements were performed with the field applied along the $a$, $b$, and $c^*$ crystallographic directions. Prior to measuring, the sample was oriented with a Laue camera. Samples were fixed on a quartz sample holder with GE-7031 varnish. High field magnetization measurements were conducted in a pulsed field up to 600 kOe. The pulsed-field measurements were performed with the $H \parallel b$ and $H \perp b$ orientations. Pulsed-field extraction magnetometer measurements provide absolute values of the change in magnetization, but must be calibrated.\cite{Goddard_2008} Therefore, we scaled the pulsed-field data to the corresponding low-field data between 50 and 70 Oe, where the where the pulsed-field $H \perp b$ results were scaled to the low-field $H \parallel c^*$ data (see Fig. \ref{fig:magnetic}).

Temperature-dependent specific heat measurements were carried out in a Quantum Design DynaCool PPMS using the relaxation technique as implemented in the heat capacity option, using heat pulses corresponding to a 2\% temperature rise, and fitting the temperature relaxation with a two-$\tau$ model. Measurements were performed under zero applied magnetic field. The sample was adhered to the platform with a small amount of Apiezon N Grease, and prior to measuring the sample, a background addenda was measured with the platform and grease only.

\textit{Results and Discussion.} Powder X-ray diffraction patterns obtained on the growth products are shown in Fig. S2 and could not be indexed by any binaries containing Y and Co or Ag. Using EDS, we found that these crystals are a ternary compound with chemical formula Y$_4$Co$_3$Ag, see Fig. S1. The refinement of our single crystal X-ray diffraction data shows that Y$_4$Co$_3$Ag adopts a new structure type with monoclinic symmetry, space group $I2/m$ (n. 12). The crystal structure is illustrated in Fig. \ref{fig:structure}, and Table \ref{tab:struc} lists refined atomic positions and thermal displacement parameters. Additional refinement details are given in Table S3.


\begin{table}[b]
    \centering
    \begin{ruledtabular}
        \caption{Fractional atomic coordinates and isotropic thermal displacement parameters of Y$_4$Co$_3$Ag determined by single crystal X-ray diffraction. The space group is $I2/m$ (n. 12). with lattice parameters of $a = 11.6716(5)$\,\AA, $b=3.9197(2)\,$\AA, $c=15.6697(6)$\,\AA, and $\beta=102.698(5)^\circ$. Refinement uncertainties are given in parentheses.}
\begin{tabular}{llllll}
Atom & Wyckoff & $x$        & $y$ & $z$        & $U_{iso}$ (\AA$^2$) \\ \hline
Y1   & $4i$    & 0.64323(2) & 0   & 0.46899(2) & 0.00965(5)                         \\
Y2   & $4i$    & 0.45504(2) & 0.5 & 0.72566(2) & 0.01019(5)                         \\
Y3   & $4i$    & 0.78875(2) & 0.5 & 0.66564(2) & 0.01108(5)                         \\
Y4   & $4i$    & 0.37191(2) & 1   & 0.90071(2) & 0.01161(5)                         \\
Co1  & $4i$    & 0.54855(3) & 0.5 & 0.57387(2)    & 0.01158(7)                         \\
Co2  & $4i$    & 0.59816(3) & 0   & 0.64834(2)    & 0.01135(7)                         \\
Co3  & $4i$    & 0.23566(4) & 0.5 & 0.78665(3) & 0.01338(7)                         \\
Ag   & $4i$    & 0.58096(2) & 0.5 & 0.93695(2) & 0.01501(5)                        
\end{tabular}
\label{tab:struc}
    \end{ruledtabular}
\end{table}

In Y$_4$Co$_3$Ag, the most striking features of the crystal structure are the zigzag and nearly hexagonal chains of Co atoms that run down the $b$-axis, occupying the channels formed by the Y atoms. Within the zigzag chains, shown in Fig. \ref{fig:structure}(c), the distance between the two nearest Co ions is $d_{\text{Co3-Co3}}$ = 2.3327(6) \AA. As highlighted in Fig. \ref{fig:structure}(b), there are two distinct Co-Co bond lengths within the hexagonal chains: $d_{\text{Co1-Co1}} = 2.3432(6)$ \AA\, and $d_{\text{Co1-Co2}} = 2.2893(6)$ \AA, which produces a slightly distorted hexagonal motif with $\gamma_1 = 121.12(1)^\circ$ and $\gamma_2 = 117.76(2)^\circ$. The minimum interchain Co-Co distance is $\sim4.2$\,\AA\,, nearly twice the $d_{\text{Co-Co}}$ values within the chains, which suggests a possible quasi-one dimensional behavior associated with the Co sublattices. Both Y and Ag atoms form linear chains extending down the $b$-axis, with much longer interchain distances of 3.9197(3)\,\AA. In agreement with the antagonistic pair hypothesis, the Y chains arrange to form small and large channels that fully sheath the respective zigzag and hexagonal Co substructures, preventing Co--Ag nearest-neighbor contacts. 

Co and Ag are among the most immiscible, or antagonistic, elements, with little/no solid solubility and a substantial immiscibility dome for liquid phases above 1000\,$^\circ$C. There are only a few known site-ordered ternary phases containing Co and Ag, and until Y$_4$Co$_3$Ag, all of these contained the most electronegative O or F atoms \cite{berthelot2011first,jezierski2024novel}. On the other hand, Y readily forms multiple binary compounds with both Co and Ag, suggesting chemical compatibility with each of the immiscible elements and, indeed, we find that low-melting Y--Co compositions can be used to grow single crystals of the intermetallic phase Y$_4$Co$_3$Ag. This is similar to our previous work with the antagonistic pairs Co--Pb and Co--Bi, in which La--Co melts led to the discovery of the compounds La$_4$Co$_4$X (X = Pb, Bi) \cite{slade2024}. More broadly, these examples suggest that, for a given antagonistic pair \textit{A}--\textit{B}, the number of \textit{A}--\textit{C} and \textit{B}--\textit{C} binary compounds with a third element \textit{C} may correlate with the existence of stable \textit{A}--\textit{B}--\textit{C} ternary phases. We believe this strategy is likely a productive direction to explore in future works. 

Figure \ref{fig:properties} presents the temperature dependence of the magnetic susceptibility $\chi = M/H$, resistivity $\rho$, and specific heat $C_p$ of Y$_4$Co$_3$Ag, with all measurements revealing a phase transition near 15\,K. The anisotropic magnetic susceptibility shown in Fig. \ref{fig:properties}(a), as well as the $M(H)$ data in Fig. \ref{fig:magnetic}, suggest that Y$_4$Co$_3$Ag orders antiferromagnetically at low temperatures. The susceptibility first increases on cooling, reaching a maximum near 15\,K, characteristic of antiferromagnetic ordering. As shown in Fig. S3(c) in the SI, the derivative of the temperature-susceptibility product \cite{fisher1962relation}, d($\chi T$)/d$T$, yields the Néel temperature $T_N$ = 14.9 K. For all three field orientations, the peak in $\chi$ is relatively sharp at the transition temperature. Below the maximum, $\chi$ decreases sharply for $H \parallel a$, but is only subtly reduced for both $H\parallel b$ and $H\parallel c^*$ orientations, suggesting that, within the antiferromagnetic state, the ordered moments may be aligned primarily along the crystallographic $a$ axis. We emphasize that, given the low symmetry of the Y$_4$Co$_3$Ag structure, \textit{H} $\parallel$ \textit{a} may only show the largest projection of the ordered moments between the three measured orientations. Neutron diffraction is ultimately needed to precisely determine the easy direction and magnetic structure of this compound.

Above the transition, $\chi$ increases in a Curie-Weiss (CW) manner and can be described by
\begin{equation}
    \chi =\frac{C}{T-\Theta} + \chi_0
\end{equation}
where $C$ is the Curie constant, $\Theta$ is the Weiss temperature, and $\chi_0$ is a $T$-independent contribution. The effective moment is estimated by $\mu_{eff}=\sqrt{8C}$. The parameters obtained from the CW fits, shown in the inset of Fig. \ref{fig:properties}(a), are summarized in Table \ref{tab:mag}. Only a weak anisotropy is observed, with all field orientations yielding comparable CW parameters. The negative Weiss temperatures are consistent with antiferromagnetic coupling between Co moments and, notably, the $\Theta$ magnitudes ($\approx$ 120 K) are much larger than the ordering temperature ($\approx$ 15 K). This suggests suppressed magnetic order compared to the strength of the magnetic exchange interactions in Y$_4$Co$_3$Ag. Furthermore, we estimate $\mu_{eff}=$1.4\,$\mu_B$/Co, somewhat smaller than the 1.73\,$\mu_B$ expected for the low-spin Co$^{2+}$ ($S=1/2$) and much smaller than the 3.87 $\mu_B$ expected for the high-spin configuration. We note that consistently large $\Theta$/$T_N$ values and comparable $\mu_{eff}$ were observed across multiple measurements with different field orientations and in samples from different batches. 

The resistivity of Y$_4$Co$_3$Ag decreases on cooling, see Fig. \ref{fig:properties}(b), with a clear kink followed by a drop at the transition temperature. The observed kink is a typical feature of magnetic ordering, marking the loss of spin-disorder scattering as the samples enter a magnetically ordered state. The resistivity drop is much stronger when the current is applied parallel to the $b$-axis, though is still observed for $I\perp b$. As seen in the inset of Fig. \ref{fig:properties}(a), the residual resistivity ratios, $\rho_{300}/\rho_{1.8}$, are relatively high, around 7--8, indicating good crystal quality. Above $T_N$, the temperature dependence deviates from the expectations for a normal metal, with $\rho$ showing a pronounced bulge between $\approx$ 50--200 K, which is also most prominent for \textit{I} $\parallel$ $\textit{b}$, see the inset of Fig. \ref{fig:properties}(b). Although we do not yet fully understand the origin of this behavior, it may be associated with electronic scattering by magnetic fluctuations that persists well above the ordering temperature \cite{ueda1977electrical}.

\begin{table}[b]
    \centering
    \begin{ruledtabular}
        \caption{Parameters obtained from Curie-Weiss fits to the high-temperature ($T>120$\,K) paramagnetic susceptibility of Y$_4$Co$_3$Ag. $\chi_0$ is reported in units of $10^{-4}$ emu/Oe/Co. Uncertainties in $\Theta$ are given in parentheses. Uncertainties of $\mu_{eff}$ and $\chi_0$ are on the order 10$^{-5}$ of the respective value.}
\begin{tabular}{lccc}
Field direction & $\Theta$ (K) & $\mu_{\mathrm{eff}}$ ($\mu_B$/Co) & $\chi_0$ \\
\hline
$H\parallel b$    & $-125(4)$ & 1.40 & 2.4 \\
$H\parallel c^*$    & $-130(4)$ & 1.46 & 3.1 \\
$H\parallel a$  & $-107(1)$ & 1.43 & 3.5 \\              \end{tabular}
\label{tab:mag}
    \end{ruledtabular}
\end{table}

Specific heat, shown in Fig. \ref{fig:properties}(c), provides evidence that the features at $\approx$15\,K observed in transport and magnetization measurements correspond to a bulk phase transition. The peak value at the $C_p$ anomaly occurs at $T_N = 14.9$\,K, which is in excellent agreement with transition temperatures estimated from the resistivity and $\chi T$ derivatives, see Fig. S3 in the SI. Without a non-magnetic reference to subtract the phonon contribution to $C_p$, we cannot quantitatively determine the entropy loss associated with magnetic ordering. However, various approaches to estimate the non-magnetic contribution to $C_p$, i.e., using a cubic spline and fits to the Debye model based on $\gamma T$ + $\beta T^3$, see inset of Fig. \ref{fig:properties}(c), all indicate a small entropy loss of $\sim$ 0.1\textit{R}ln(2) or less. Such a small entropy loss associated with magnetic ordering is observed in other small moment, itinerant magnetic materials \cite{CLINTON197573,takeuchi1979low,xiang2021avoided,svanidze2015itinerant}, supporting the picture that Y$_4$Co$_3$Ag has a small ordered moment and/or that strong fluctuations persist well above $T_N$.


Figure \ref{fig:magnetic} presents field-dependent magnetization isotherms $M(H)$ for Y$_4$Co$_3$Ag. The anisotropic response at 1.8\,K, Fig. \ref{fig:magnetic}(a), reveals that the low-field $M(H)$ is smallest when the magnetic field is applied along $a$. The $H\parallel a$ isotherm shows a field-induced transition near 40\,kOe that is complete by 60\,kOe, above which the magnetization is approximately linear and remains unsaturated up to 70 kOe. For $H\parallel b$ and $H\parallel c^*$, $M(H)$ is effectively linear up to at least 70\,kOe. As the magnetization remains unsaturated up to 70 kOe, we performed pulsed field measurements up to 600 kOe at 500\,mK. Fig. \ref{fig:magnetic}(b) presents the data collected with $H \parallel b$ and $H \perp b$ (chosen for convenient sample alignment). In both orientations, $M(H)$ increases linearly up to $\sim$300\,kOe, where a kink is observed and above which the magnetization increases more gradually, trending toward 0.14 $\mu_B$ at 600 kOe. Despite remaining unsaturated at 600 kOe, the modest values of \textit{M(H)} found at the highest field measurements are consistent with Y$_4$Co$_3$Ag having a small ordered moment. If we consider 0.2 $\mu_B$/Co as an upper-bound on the saturated moment, this produces an estimated Rhodes-Wohlfarth ratio  $q_c/q_s$ of $\sim 4.0$ \cite{rhodes1963effective}. As the real ordered moment is likely somewhat smaller, these results place Y$_4$Co$_3$Ag squarely in the itinerant magnetism regime \cite{takahashi2013spin}.

Taken together, our measurements show that Y$_4$Co$_3$Ag orders antiferromagnetically at $T_N = 14.9$\,K, with an effective moment of 1.4 $\mu_B$, a small specific heat anomaly (entropy loss), and low magnetization under 0.2 $\mu_B$ at 500 mK. All these features suggest that Y$_4$Co$_3$Ag carries a relatively small ordered moment. Notably, Curie-Weiss analysis of the paramagnetic magnetization indicates a relatively high ratio of $\Theta$/$T_N$ $\sim$ 9, consistent with relatively strong exchange interactions compared to the scale of the ordering temperature. Traditionally, enhanced $\Theta$/$T_N$ $\approx$ 10 is recognized as a signature of strong magnetic frustration \cite{moessner2006geometrical}. Here, the crystal structure of Y$_4$Co$_3$Ag does not suggest geometrical frustration. Alternatively, considering the modest $\Theta$/$T_N$ $\approx$ 1--2 observed in other itinerant magnets with more unambiguously 3D crystal structures \cite{gardner1968magnetization,BLYTHE1966144,svanidze2015itinerant,PhysRevB.105.014412}, it is possible that the one-dimensional Co chains lead to strong magnetic fluctuations that suppress long-range order relative to the Weiss temperature. The presence of strong fluctuations is roughly consistent with the substantial bulge observed in resistivity, suggesting strong magnetic scattering that persists well above the ordering temperature.

Nevertheless, a characteristic of low-dimensional interactions and ordering, typically observed in insulators, is a broad maximum in the temperature-dependent susceptibility and deviations from Curie-Weiss-like behavior immediately above their ordering temperature \cite{vasiliev2018milestones}. In contrast, Y$_4$Co$_3$Ag shows a sharp peak in $\chi$ at $T_N$ and CW-like behavior in the paramagnetic state, both features commonly associated with 3D magnetic interactions/ordering. In this context, Y$_4$Co$_3$Ag is in many ways similar to La$_5$Co$_2$Ge$_3$ \cite{saunders2020exceedingly, xiang2021avoided}, another small moment, Co-based itinerant magnetic system which orders around 4\,K with an inferred ordered moment of $\sim0.1\,\mu_B$ and Rhodes-Wohlfarth ratio $q_c/q_s = 4.9$, higher than several well-established itinerant ferromagnetic systems. Similarly to Y$_4$Co$_3$Ag, La$_5$Co$_2$Ge$_3$ also exhibits sharp features in $\chi(T)$, $\rho(T)$ and $C_p$ associated with the onset of long-range magnetic order and has a pronounced bulge in the temperature dependence of the resistivity, deviating from the expectation for a normal metal. Likewise, La$_5$Co$_2$Ge$_3$ also features Co atoms in a quasi-low dimensional arrangement, with Co-Co dimers that form ladder-like chains along the \textit{b}-axis. However, La$_5$Co$_2$Ge$_3$ orders ferromagnetically and has a smaller ratio of $\Theta/T_N$ $\approx$ 3. Ultimately, Y$_4$Co$_3$Ag is an interesting system that demands further theoretical and experimental investigation to determine whether its electronic structure and magnetic interactions bear the fingerprints of low-dimensionality. Furthermore, it offers a chance to explore potential emergent phenomena upon suppressing the magnetic ordering with pressure, doping, or magnetic field.

\textit{Conclusion.} We have outlined the discovery of Y$_4$Co$_3$Ag, the first intermetallic compound containing the antagonistic pair Co--Ag. Due to the strong Co--Ag immiscibility, Y$_4$Co$_3$Ag has a remarkable crystal structure, with zigzag and hexagonal Co chains running along the crystallographic \textit{b}-axis. These chains are separated from the Ag atoms by channels formed by Y. Transport, magnetic, and thermodynamic measurements reveal that Y$_4$Co$_3$Ag orders at $T_N = 14.9$\,K, with Co ions coupled antiferromagnetically and carrying a small ordered moment, likely under 0.2 $\mu_B$/Co. The ordering temperature is significantly lower than the Weiss temperature, $\Theta/T_N\sim9$. This, together with an anomalous temperature dependence of resistivity and a small entropy loss from magnetic ordering, all indicate that strong magnetic fluctuations may persist well above $T_N$ in Y$_4$Co$_3$Ag. As such, Y$_4$Co$_3$Ag is a new itinerant, small moment, magnetic system and, given the low-dimensionality of Co chains, may be an attractive platform to explore low-dimensional magnetic interactions associated with itinerant electrons. More broadly, this work is a successful demonstration of the antagonistic pair design principle for discovering intermetallic phases with magnetic ions embedded in low dimensional motifs. 

Work at Ames National Laboratory was supported by the U.S. Department of Energy (DOE), Basic Energy Sciences, Division of Materials Sciences \& Engineering, under Contract No. DE-AC02-07CH11358. RFSP's one-year visit to Ames Laboratory and Iowa State University was supported by the São Paulo Research Foundation (FAPESP), Brasil, Process Number 2024/08497-6. RFSP also acknowledges support from FAPESP under Process Number 2021/01004-6. SLM acknowledges support from FAPESP under Process Number 2023/10775-1. A portion of this work was performed at the National High Magnetic Field Laboratory (NHMFL), which is supported by NSF Cooperative Agreement Nos. DMR-1644779 and DMR-2128556, the state of Florida, and the Department of Energy (DOE). J.S. acknowledges support from the DOE BES program “Science at 100 T”. We would like to thank A. Sapkota for assistance with single crystal orientation using the Laue camera.

\textit{Data Availability.} The data that support the findings of this study will be made available in DataShare, an open-access repository at Iowa State University.

*corresponding authors' email: slade@ameslab.gov, rafaelafelixp@usp.br

\bibliography{bib}

\end{document}


\title{Supporting Information \\
\vskip 0.5cm 
Itinerant antiferromagnetism in the antagonistic pair compound Y$_4$Co$_3$Ag}

\author{Rafaela F. S. Penacchio}
\affiliation{Ames National Laboratory, US DOE, Iowa State University, Ames, IA, USA}
\affiliation{Department of Physics and Astronomy, Iowa State University, Ames, Iowa 50011, USA}
\affiliation{Institute of Physics, University of S{\~{a}}o Paulo, S{\~{a}}o Paulo, SP, Brazil}

\author{Nao Furukawa}
\affiliation{Ames National Laboratory, US DOE, Iowa State University, Ames, IA, USA}
\affiliation{Department of Physics and Astronomy, Iowa State University, Ames, Iowa 50011, USA}

\author{Joanna M. Blawat}
\affiliation{National High Magnetic Field Laboratory,
Los Alamos National Laboratory, Los Alamos, NM, USA.}

\author{John Singleton}
\affiliation{National High Magnetic Field Laboratory,
Los Alamos National Laboratory, Los Alamos, NM, USA.}

\author{Zhouqi Li}
\affiliation{Ames National Laboratory, US DOE, Iowa State University, Ames, IA, USA}
\affiliation{Department of Physics and Astronomy, Iowa State University, Ames, Iowa 50011, USA}

\author{Raquel~A. Ribeiro}
\affiliation{Ames National Laboratory, US DOE, Iowa State University, Ames, IA, USA}
\affiliation{Department of Physics and Astronomy, Iowa State University, Ames, Iowa 50011, USA}


\author{S{\'{e}}rgio L. Morelh{\~{a}}o}
\affiliation{Institute of Physics, University of S{\~{a}}o Paulo, S{\~{a}}o Paulo, SP, Brazil}

\author{Sergey L. Bud’ko}
\affiliation{Ames National Laboratory, US DOE, Iowa State University, Ames, IA, USA}
\affiliation{Department of Physics and Astronomy, Iowa State University, Ames, Iowa 50011, USA}

\author{Paul C. Canfield}
\affiliation{Ames National Laboratory, US DOE, Iowa State University, Ames, IA, USA}
\affiliation{Department of Physics and Astronomy, Iowa State University, Ames, Iowa 50011, USA}

\author{Tyler J. Slade}
\affiliation{Ames National Laboratory, US DOE, Iowa State University, Ames, IA, USA}
\affiliation{Department of Physics and Astronomy, Iowa State University, Ames, Iowa 50011, USA}
\renewcommand{\thefigure}{S\arabic{figure}}
\setcounter{figure}{0}

\renewcommand{\thetable}{S\arabic{table}}
\setcounter{table}{0}

\maketitle
 
\clearpage
\onecolumngrid

\newpage

\section{Comparison of Y$_4$Co$_3$Ag single crystals grown in Al$_2$O$_3$ and Ta crucibles}

As discussed in the \textit{Methods} section of the main text, we succeeded in growing Y$_4$Co$_3$Ag single crystals in reactions contained either in Ta or Al$_2$O$_3$ crucibles. In the latter, the elements were also weighted in a molar ratio of Y$_{60}$Co$_{40}$Ag$_{10}$ and then arc-melted on a water-cooled copper hearth using Zr pellets as an oxygen getter. The melted button was flipped and thoroughly remelted five times to ensure a homogeneous melt. The mass loss after arc-melting was smaller than 0.1\,mg. The melted button was loaded into a 2\,ml Al$_2$O$_3$ fritted Canfield crucible set (CCS) \cite{canfield2001, LSP}, and the packed CCS was flame-sealed into a fused silica ampule that was backfilled with $1/6$ atm Ar. The ampule was heated in a box furnace to 900$\,^\circ$C. After dwelling at 900\,$^\circ$C for 6\,h, the furnace was cooled to 760\,$^\circ$C over 125\,h. The decanting step is analogous to that described in the main text. After cooling to room temperature, the ampule and CCS were opened to reveal a cluster of rod-like crystals.

\begin{figure*}[ht]
    \centering
    \includegraphics[scale=1.1]{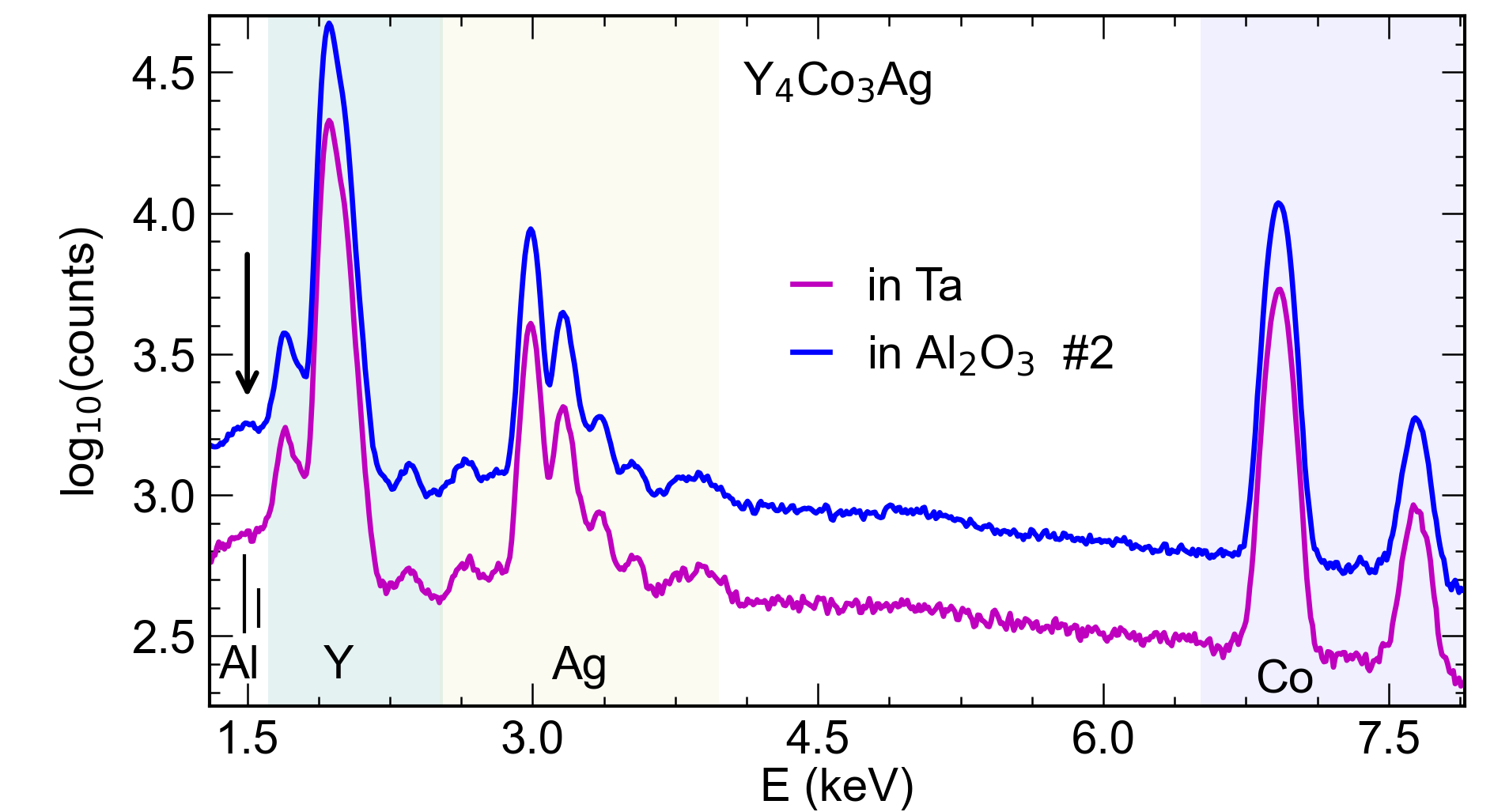}
    \caption{EDS spectra obtained on Y$_4$Co$_3$Ag single crystals grown in Al$_2$O$_3$ (batch \#2, see Table \ref{tab:prop}) and Ta crucibles. Green, yellow, and purple shaded areas indicate the position of Y-L, Co-K, and Ag-L spectral lines, respectively. The K-lines of Al are also shown and indicated by the black arrow.}
    \label{fig:eds}
\end{figure*}

When using Al$_2$O$_3$ crucibles, the Y-rich starting melt may undergo a thermite-like reaction with the crucible, potentially introducing Al metal into the melt that could be incorporated into the growth products. To investigate this possibility, we performed EDS on Y$_4$Co$_3$Ag single crystals obtained by both methods. As shown in Fig. \ref{fig:eds}(a), the spectra obtained on samples grown in Al$_2$O$_3$ crucibles exhibit a very small, but detectable, amount of Al contamination, whereas no Al is detected in crystals grown in Ta tubes. The average composition of both samples, presented in Table \ref{tab:eds}, is consistent with the 4:3:1 stoichiometry, but crystals grown in Al$_2$O$_3$ show a maximum incorporation of $\sim0.1\%$ Al.

\begin{table}[b]
    \centering
        \caption{EDS data (atomic \%) for Y$_4$Co$_3$Ag samples grown in Ta and Al$_2$O$_3$ crucibles. The data was collected on three different spots (\#1 to \#3) for each sample. Average compositions are given in the last row. Uncertainties in parenthesis.}
\begin{tabular}{lllll|llll}
       & \multicolumn{4}{c}{in Ta} & \multicolumn{4}{c}{in Al$_2$O$_3$} \\ \hline
       & Y    & Co   & Ag   & Al   & Y       & Co     & Ag     & Al     \\ \hline
\# 1 & 51.7 & 35.6 & 12.7 & 0.01 & 51.5    & 35.8   & 12.5   & 0.12   \\
\# 2  & 51.7 & 35.5 & 12.8 & 0.04 & 51.5    & 35.9   & 12.6   & 0.11   \\
\# 3   & 51.8 & 35.4 & 12.8 & 0    & 51.5    & 35.6   & 12.8   & 0.11  \\ \hline
Comp.      & {4.0(3)}    &  {2.8(2)}  &  {1} &  {0.000(1)} &  {4.1(5)} & {2.8(4)} & {1} & {0.010(1)} \\ \hline
\end{tabular}
\label{tab:eds}
\end{table}


\begin{figure*}[ht]
    \centering
    \includegraphics[scale=.9]{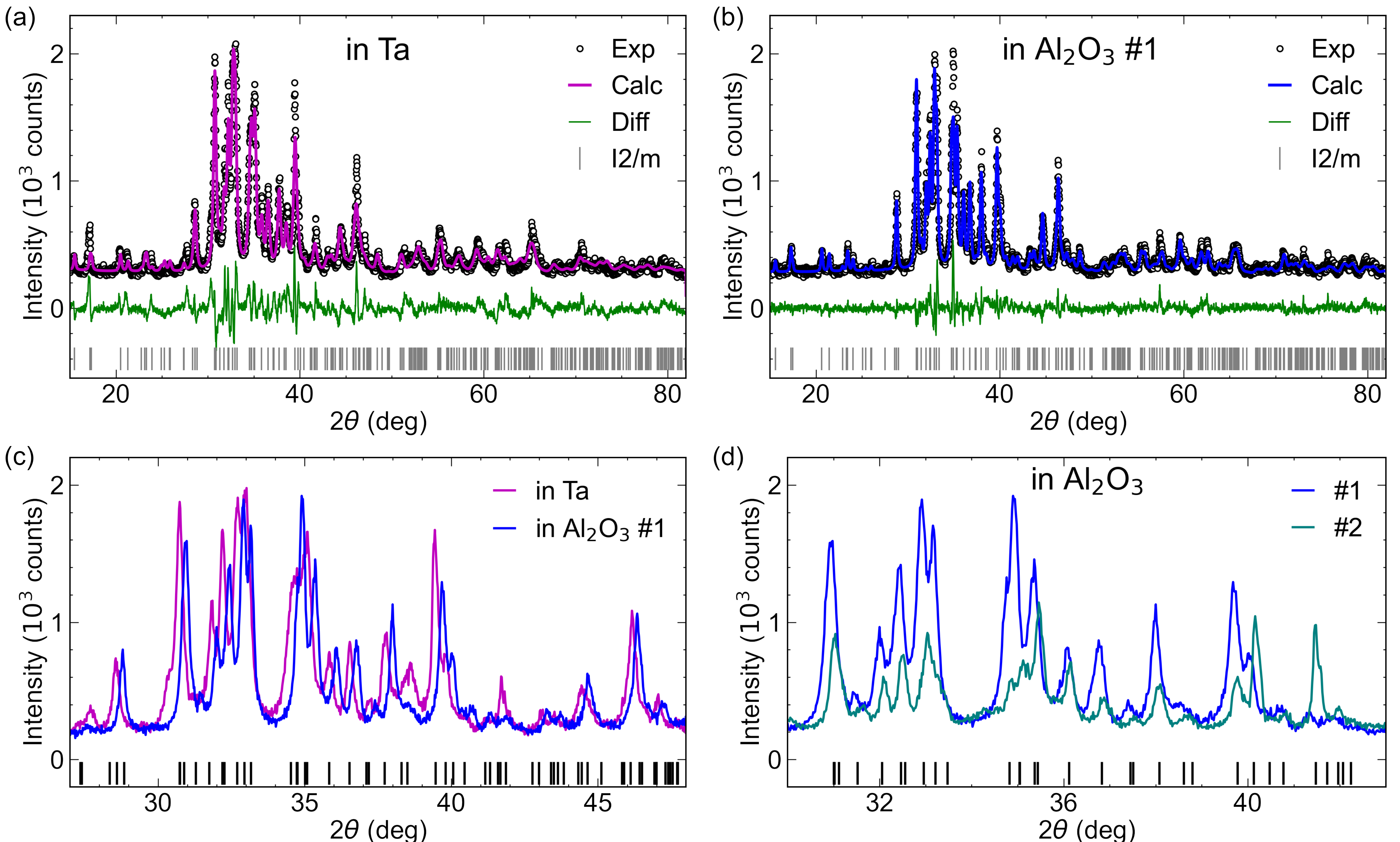}
    \caption{Powder X-ray diffraction patterns and Rietveld refinements for Y$_4$Co$_3$Ag grown in (a) Ta and (b) Al$_2$O$_3$ crucibles. $R_{wp}=9.8$ $\%$ and 15.4 $\%$ for the refinements in (a) and (b) respectively. (c) Close-up view of the patterns in (a) and (b) around $2\theta\sim38^\circ$, emphasizing the lattice parameter difference between these samples. Black marks indicate the positions of Bragg peaks for the sample grown in Ta. (d) Comparison of powder patterns obtained on two different batches, \#1 and \#2, of crystals grown in Al$_2$O$_3$ crucibles. The black dashes in (c) and (d) mark the expected peak positions for Y$_4$Co$_3$Ag based on the refinement of Al$_2$O$_3$-grown sample \#1. The Al$_2$O$_3$-based data in (b) and (c) corresponds to batch \#1.}
    \label{fig:pxrd}
\end{figure*}

\begin{table}[hb!]
    \centering
    \setlength{\tabcolsep}{4pt}
    \begin{ruledtabular}
    \caption{Comparison of the lattice parameters, ordering temperatures, and magnetic properties of Y$_4$Co$_3$Ag single crystals grown in Ta and Al$_2$O$_3$ crucibles. In the latter, data corresponding to single crystals from two different batches, \#1 and \#2, are shown. Y$_4$Co$_3$Ag adopts a monoclinic structure (space group $I2/m$), with lattice parameters $a$, $b$, $c$, $\beta$, and unit cell volume $V$. Residual-resistance-rations, $RRR=R(300\,\mathrm{K})/R(1.8\,\mathrm{K})$, are also shown. Néel temperatures inferred from local maxima in resistance and $\chi T$ derivatives, shown in Figs. \ref{fig:alumina}(a, c), are indicated by $T_R$ and $T_\chi$, respectively. Resistances were measured with current applied parallel to $b$-axis, and magnetic susceptibility with $H\parallel b$ (for convenient sample orientation). Weiss temperature $\Theta$, effective moment $\mu_{eff}$ ($\mu_B$/Co), and temperature-independent contribution to the magnetic susceptibility $\chi_0$ ($10^{-4}$\,emu/Oe/Co) were obtained from the Curie-Weiss fits shown in Fig. \ref{fig:alumina}(b). Uncertainties of $\mu_{eff}$ and $\chi_0$ (not shown) are 10$^{3}$ smaller than the corresponding values.}
\begin{tabular}{clllllllllll}
Y$_4$Co$_3$Ag       & \multicolumn{1}{c}{$a$ (\AA)} & \multicolumn{1}{c}{$b$ (\AA)} & \multicolumn{1}{c}{$c$ (\AA)} & \multicolumn{1}{c}{$\beta$ ($^\circ$)} & \multicolumn{1}{c}{$V$ (\AA$^3$)} & $RRR$ &\multicolumn{1}{c}{$T_R$ (K)} & \multicolumn{1}{c}{$T_{\chi}$ (K)} & $\Theta$ (K) & $\mu_{eff}$ & $\chi_0$ \\ \hline
Ta               & 11.712 (3)                                   & 3.9333 (5)                                   & 15.748 (5)                                   & 102.69 (1)                             & 707.8 (1)     &     7.9                              & 14.6                                       & 14.9 & -125 (4)     & 1.4                      & 2.4                            \\
Al$_2$O$_3$ \#1 & 11.653 (1)                                   & 3.9149 (2)                                   & 15.635 (2)                                   & 102.710 (5)                            & 695.8 (1)     &     5.3                             & 14.3                                       & 14.8                                              & -80 (1)      & 1.2                      & 2.8                            \\
Al$_2$O$_3$ \#2 & 11.628 (2)                                   & 3.9016 (5)                                   & 15.596 (5)                                   & 102.68 (1)                             & 690.3 (1)     &     4.9                             & 12.5                                       & 13.0                                              & -67 (19)     & 1.0                      & 4.8                            
\end{tabular}
\label{tab:prop}
    \end{ruledtabular}
\end{table}

Single crystal X-ray diffraction data do not provide any evidence for disordered/interstitial sites on samples grown in Al$_2$O$_3$. Furthermore, as seen in the Table \ref{tab:refinement}, the refinement statistics are excellent and are nearly identical with those from Ta-grown samples, which prevent us from speculating on how Al is being incorporated into the structure. Overall, these results are consistent with the very small $\sim$ 0.1 $\%$ fraction of Al detected by EDS. 

Notably, the Al contamination is clearly seen through powder X-ray diffraction. Though EDS supports only a small $\sim$ 0.1 $\%$ Al incorporation, we observe a shift of all reflections to higher angles, implying a reduced volume of samples grown in Al$_2$O$_3$, see Fig. \ref{fig:pxrd}(c). This volume compression is likely associated with substitution of smaller Al atoms into the Y$_4$Co$_3$Ag structure. Changes in the residual-resistance-ratios, see following discussion, is also consistent with Al incorporation into the crystal structure. Lattice parameters obtained from Rietveld refinements are given in Table \ref{tab:prop}. Overall, the Al contamination results in an almost uniform compression of $a$, $b$, and $c$ parameters, with $\beta$ remaining nearly unchanged, thus reducing the volume by 2 to 3\%. Fig. \ref{fig:pxrd}(d) emphasizes that, among samples grown in Al$_2$O$_3$ crucibles, we observe a variation in the peak shifts and relative peak intensities in our powder patterns, suggesting that the degree of Al incorporation is not fully reproducible. Given that the Al is introduced by partial reaction between the starting elements and the crucible, batch-to-batch inconsistency in the degree of Al substitution is not surprising.

\begin{figure*}[ht]
    \centering
    \includegraphics[scale=.9]{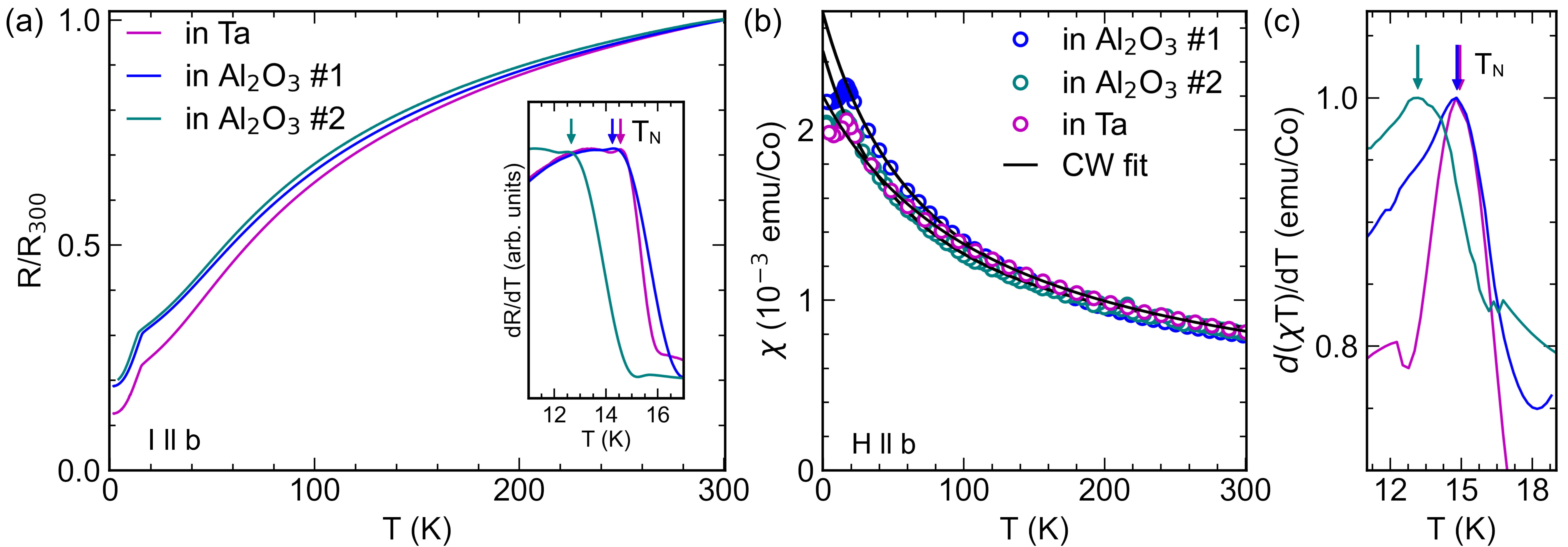}
    \caption{Physical properties of Y$_4$Co$_3$Ag single crystals grown in Ta and Al$_2$O$_3$ crucibles. Data from two different batches of crystals grown in Al$_2$O$_3$, \#1 and \#2, are shown. (a) Temperature-dependent resistance normalized to its value at 300\,K, $R/R_{300}$. The low-temperature dependence of resistance derivative, $dR/dT$, is shown in the inset. The arrows indicate the local maximum in $dR/dT$ near the magnetic ordering transition, which was used to define the Néel temperature $T_N$. (b) Temperature dependence of the magnetic susceptibility $\chi$. All measurements were field-cooled with a $H =10$\,kOe field applied parallel to the $b$-axis. Solid lines are the corresponding modified Curie-Weiss fits discussed in the text. (c) Derivative of $\chi T$, i.e. $d(\chi T)/dT$, around the ordering temperature. Arrows indicate $T_N$, as defined by the maximum of $d(\chi T)/dT$.}
    \label{fig:alumina}
\end{figure*}

Fig. \ref{fig:alumina} compares the temperature-dependent resistance and magnetic susceptibility of Y$_4$Co$_3$Ag crystals grown in Ta with samples from two different batches that used Al$_2$O$_3$ crucibles. Overall, the data is qualitatively similar across the different batches; however, the second Al$_2$O$_3$-based growth shows a slightly suppressed ordering temperature in both the resistance and $\chi T$ derivatives, as displayed in the inset of Fig. \ref{fig:alumina}(a) and Fig. \ref{fig:alumina}(c). This likely suggests that Al incorporation can suppress the ordering temperature, and that the second Al$_2$O$_3$-grown sample contains a higher Al fraction. The inferred Néel temperatures are listed in Table \ref{tab:prop} (columns 7 and 8). The residual-resistance-ratios, $RRR=R(300\,\mathrm{K})/R(1.8\,\mathrm{K})$, are $\sim$8 and 5 for samples grown in Ta and Al$_2$O$_3$ (batches \#1 and \#2), respectively. The smaller, but still reasonable, values observed for Al-contaminated samples are consistent with a small degree of disorder. As shown in Fig. \ref{fig:alumina}(b), the magnetic susceptibilities of all samples are Curie-Weiss-like above $T_N$. The fitted parameters are listed in the Table \ref{tab:prop} (columns 9--11). We found that $\mu_{eff}$ is slightly smaller in the Al-contaminated samples, and that $\Theta$ is more substantially reduced. Both results are consistent with Al incorporation suppressing the magnetic interactions and long range order. The observed trends suggest that even a small amount of Al incorporation can, in principle, change the magnetic properties of Y$_4$Co$_3$Ag; thus motivating a future work on controlled doping studies of this compound.

\clearpage
\section{Magnetization isotherms at different temperatures}

\begin{figure*}[ht]
    \centering
    \includegraphics[width=\linewidth]{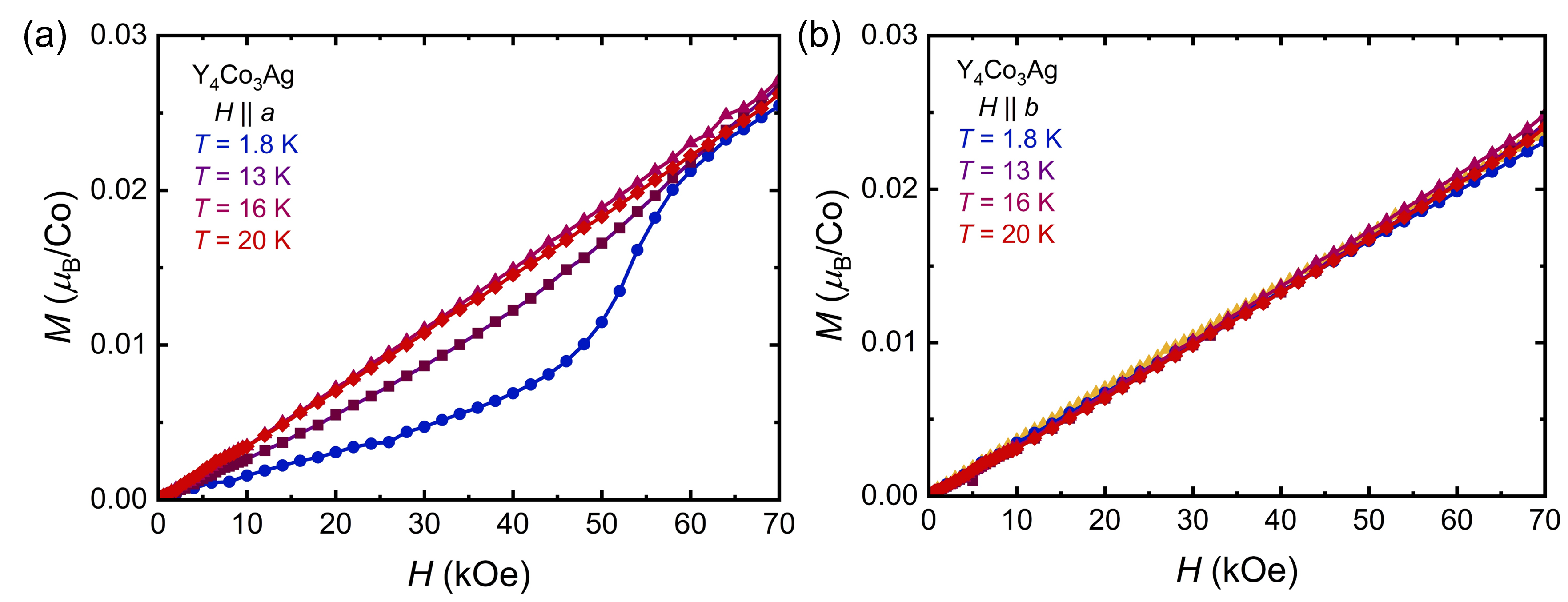}
    \caption{Magnetization isotherms of Y$_4$Co$_3$Ag measured at different temperatures in the (a) $H \parallel a$ orientation and (b) $H \parallel b$ orientation.}
    \label{MH_Ts}
\end{figure*}

In addition to the \textit{T} = 1.8 K magnetization isotherms shown in Figure 3 of the main text, we also measured \textit{M}(\textit{H}) at several temperatures above and below the Néel temperature (14.9 K). The results are presented in Figure S4 for fields oriented along the $a$-axis (Figure S4(a)) and along the $b$-axis (Figure S4(b)). Above $T_{\text{N}}$, the \textit{T} = 20 K and \textit {T} = 20 K magnetization curves are linear in field up to at least 70 kOe for both orientations. For \textit{T} = 13 K, the $H \parallel a$ isotherm shows subtly sub-linear behavior, likely reflecting substantial broadening of the $\approx$ 50 kOe metamagnetic transition that is much more sharply observed at 1.8 K in this orientation. For $H \parallel b$, the \textit{M}(\textit{H}) plots are nearly invariant with temperature between 1.8--20 K.

The isotherms shown in Fig. S4(b) were collected independently from the magnetic data shown in the main text. Here, the measurements were done using a Quantum Design Magnetic Properties Measurement System (MPMS-classic) in the DC measurement mode. The sample was glued to a kel-F disc, which was held inside of a straw to conduct the measurement. Prior to measuring the sample, we measured the bare disc at the same temperatures and fields and used these results as a background subtraction.

\clearpage
\section{Crystallographic refinement details}

Table \ref{tab:refinement} provides information regarding the structural refinements of single crystal data for Y$_4$Co$_3$Ag samples grown either in Ta and Al$_2$O$_3$ crucibles.

\begin{table*}[htbp]
    \centering
    \begin{ruledtabular}
    \caption{Single crystal data and structural refinement information for Y$_4$Co$_3$Ag.}
    \begin{tabular}{lll}
                                            & \multicolumn{1}{c}{in Ta}     & \multicolumn{1}{c}{in Al$_2$O$_3$ \#1}    \\ \hline
Chemical formula                            & Y$_4$Co$_3$Ag                 & Y$_4$Co$_3$Ag                  \\
F.W (g/mol)                                 & 640.30                       & 640.30                        \\
Space group, $Z$                            & $I2/m$ (\# 12), 4                   & $I2/m$ (\# 12), 4           \\
$a$ (\AA)                                   & 11.6716 (5)                    & 11.6708 (4)                   \\
$b$ (\AA)                                   & 3.9197 (2)                    & 3.9214 (1)                   \\
$c$ (\AA)                                   & 15.6697 (6)                   & 15.6625 (5)                   \\
$\beta$ ($^\circ$)                          & 102.698 (4)                    & 102.708 (4)                   \\
Volume (\AA$^3$)                            & 699.34 (5)                   & 699.25 (4)                   \\
$\theta$ range ($^\circ$)                    & 3.128-63.95                  & 3.874-63.788                  \\
No. reflections, $R_{int}$                  & 9304, 0.0365                  & 9669, 0.0239                        \\
No. independent reflections                 & 2306                           &  2328                       \\
No. parameters                              & 50                            & 50                           \\
$R_1$, $wR_2$ [$I>2d(I)$]                   & 0.0215, 0.0443                 & 0.0177, 0.0364               \\
Goodness of fit                             & 1.056                         & 1.021                       \\
Diffraction peak and role ($e^{-}$/\AA$w^{3}$) & 1.24, -1.66                  & 0.79, -1.16  \\
CSD number & 2548209 & 
\end{tabular}
\label{tab:refinement}
    \end{ruledtabular}
\end{table*}




\newpage

\bibliography{si}